\documentclass[12pt]{aa}
\usepackage[english]{babel}
\usepackage{graphicx}

\onecolumn
\newcommand{\ab}{Astrophys. Bull. }
\newcommand{\bsao}{Bull. Spec. Astrophys.Observ.}
\newcommand{\arep}{Astron. Rep. }
\newcommand{\alet}{Astron. Let. }
\newcommand{\araa}{Ann. Rev. Astron. Astrophys. }
\newcommand{\mnras}{Mon. Not. R. Astron. Soc. }
\newcommand{\apj}{Astrophys. J. }

\newcommand{\aj}{Astron. J.}
\newcommand{\aaa}{Astron and Astrophys.}
\newcommand{\aas}{Astron and Astrophys. Suppl.}

\def\pasj{{Publ. Astron. Soc. Japan}}
\def\rmxaa{Revista Mex. Astron. Astrofis,}

\begin{document}

\title{Optical spectrum of distant OH/IR star V1648\,Aql (IRAS\,19386+0155)}
\author{V.G.~Klochkova and N.S.~Tavolzhanskaya \\
{\small \email{valenta@sao.ru}}}

\institute{Special Astrophysical Observatory RAS, Nizhnij Arkhyz,  369167 Russia}
\date{\today} 

\abstract{An optical spectrum of the star V1648\,Aql (=\,IRAS\,19386+0155) was 
obtained at the 6-meter  telescope with a spectral resolution of R$\ge60\,000$. 
Heliocentric radial velocity measured from numerous metallic absorptions is equal to 
V$_r$=10.18$\pm0.05$\,km\,s$^{-1}$ (V$_{\rm LSR}$=18.1\,km s$^{-1}$).
We determined the atmospheric, circumstellar, and interstellar
components in the profile of the Na\,I D~lines at V$_r$=9.2, $-3.4$, and 
$-12.8$\,km s$^{-1}$ respectively. The averaged over twenty identified DIBs velocity 
V$_r$(DIBs)=$-12.5\pm0.2$\,km~s$^{-1}$ coincides with the interstellar Na\,I component. 
Weak emissions with an intensity of about 10\% of the local continuum level were 
detected in the spectrum; they are identified as low-excitation metal lines. 
Their averaged position, V$_r$=8.44$\pm0.28$\,km s$^{-1}$, may point to the presence 
of a weak velocity gradient in the upper layers of the stellar atmosphere. Based on 
the spectroscopic data and taking into account the interstellar and circumstellar reddening, 
we estimated the star's luminosity M$_V\approx-5^{\rm m}$ and also obtained the lower 
estimate of distance d$\ge1.8$ kpc. Using the model atmosphere method, we determined the 
fundamental parameters and chemical abundances in the atmosphere approving the status of
a post-AGB star for V1648\,Aql.
\keywords{stars: evolution---stars: individual: V1648\,Aql---stars: AGB and post-AGB}
}
\titlerunning{\small Optical spectrum of V1648\,Aql}
\authorrunning{\small Klochkova \& Tavolzhanskaya}

\maketitle

\section{Introduction}

Based on the photometric observations in the visible and IR ranges, the authors of~[1] 
have concluded the belonging of a point infrared IRAS\,19386+0155 source to the
objects near the asymptotic giant branch (AGB). Despite a significant IR flux, there is 
no  OH maser emission in the 1612-MHz band~[2] which indicates a later post-AGB stage. 
The re-analysis of the Arecibo Observatory survey data made it possible to detect a weak 
radiation of the IRAS\,19386+0155 source in the OH maser bands~[2]. Based on the near 
IR polarimetry, Gledhill~[4] suspected a possible belonging of IRAS\,19386+0155 to bipolar nebulae.

At the post-AGB stage, far-evolved stars with initial masses within the interval of 
2$\div8M_{\sun}$ are observed. At the preceding AGB stage, these stars are observed 
as red supergiants with  effective temperature is $T_{\rm eff}\approx 3000\div 4500$\,K.
For the stars of the above masses the AGB stage is the final evolutionary 
stage with nuclear burning in their interiors~[5]. The interest to AGB stars 
and their closest descendants can be explained primarily by the fact that namely 
in the interiors of these stars, being at a short evolutionary stage,  physical conditions 
are created for the nuclear synthesis and ejection of the products of nuclear
reactions into the stellar atmosphere and further into the circumstellar and interstellar 
medium. Thus, AGB stars of initial masses smaller than 3$\div4M_{\sun}$ are the main sources of
heavy metals (over 50\% of all the elements heavier than iron) synthesized as a result of 
the $s$-process, the essence of which is a slow (compared to $\beta$ decay) neutronization of
nuclei. A seed nucleus for a series of reactions in the $s$-process is the Fe nucleus. 
For stars with initial masses smaller than 3$\div4$M$_{\sun}$, the required neutron flux is
provided by the  $^{13}$C$(\alpha,n)^{16}$O reaction, while in the case of more massive 
stars with initial masses greater than 4$\div5 M_{\sun}$, a similar reaction takes place with the
$^{22}$Ne nuclei. These more massive AGB stars can be the sources of lithium. The evolutional 
features of  stars near the AGB and results of the current calculations of synthesis and ejection of
elements are given in the papers~[6--8].

In the visible range, the IRAS\,19386+0155 source is associated with the supergiant V1648\,Aql 
of the F5\,I spectral type~[9]. This supergiant is located outside the Galaxy plane that already 
indicates its possible belonging to the evolved stars with initial masses of 2$\div$8~$M_{\sun}$. 
To date, the star was studied by photometric methods mainly. In particular, Arkhipova et al.~[10], 
having conducted a 19 year long $UBV$ monitoring of the star, studied the light curve and revealed 
a sinusoidal brightness variability with an amplitude typical of post-AGB stars and an about 100$^d$
period. Let us notice a complex character of the long-term variability of color indices of V1648\,Aql 
detected by these authors, interpretation of which requires a further monitoring of the star. 
Later, Hrivnak et al.~[11], having added a lot of photometric data obtained by other authors
including~[10] to their own observations, confirmed the periods of brightness variability and trends in the
brightness and color index variabilities. V887\,Her, the central star of the IR source IRAS\,18095+2704, 
is close to V1648\,Aql by the set of observed parameters. Based on the two-peaked energy distribution 
and a presence of OH maser radiation, Hrivnak et al.~[12] consider IRAS\,18095+2704 as a prototype 
of O-rich protoplanetary nebulae. Lewis~[3], analyzing the properties of a broad sample of high-latitude 
OH/IR stars, also emphasized the similarity of the evolutionary state of the protoplanetary nebulae
IRAS\,18095+2704 and IRAS\,19386+0155.

Being rather faint in the visible range, the star V1648\,Aql was very rarely studied via optical 
spectroscopy. Pereira~et~al., in their paper~[13] modeled the stellar atmosphere using a high-resolution 
spectrum in a wide wavelength range and determined the fundamental parameters and
chemical abundance of the central star of this source. The authors concluded that V1648\,Aql is an 
O-rich star with an effective temperature  T$_{\rm eff}=6800\pm100$\,K, the surface gravity
$\log g$=1.4$\pm$0.2, and a low metallicity [Fe/H]$_{\sun}$\,=\,$-1.1$. They also estimated the abundances 
of a number of light metals and heavy elements in the atmosphere. An important result of
study~[13] is modeling of an unusual for post-AGB stars energy distribution in the stellar spectrum, 
which allowed the authors to conclude on a possible presence of a dust disk around the
object. Unfortunately, these authors  paid little attention to the features of the optical spectrum, 
and the paper neither has any data on the radial velocity pattern in the system at all. Thus, the need 
to continue the study of the system  IRAS\,19386+0155 is obvious.

In this paper, we present an analysis of the optical spectrum of
V1648\,Aql obtained in 2017. Section~\ref{obs}
briefly describes the methods of observations and data analysis. In
Section~\ref{results}, we present our results compared
to those published earlier, and Section~\ref{conclus}
gives the conclusions.

\section{Observations, reduction and analysis of the spectra}\label{obs}

The spectrum of V1648\,Aql was obtained on August 7, 2017 with the
NES echelle spectrograph~[14] permanently mounted at the Nasmyth focus of 
the 6-m BTA telescope of the Special Astrophysical Observatory of the Russian 
Academy of Sciences. At this date, the NES echelle spectrograph was
equipped with a $4608\times2048$ elements CCD detector with a pixel size 
of $0.0135\times0.0135$\,mm and readout noise of 1.8~e$^-$. The registered 
spectral range is $\Delta\lambda=470$--$778$~nm. In order to reduce the light loss
without any spectral resolution decrease, the NES spectrograph is equipped 
with a three-slice image slicer. Each spectral order in a 2D image of the spectrum 
is  repeated thrice, being shifted along the echelle-grating dispersion~[14]. The
spectral resolution is $\lambda/\Delta\lambda\ge60\,000$, the signal-to-noise ratio 
is $S/N>100$, varying along the echelle order from 100 to 150.

\begin{table*}[ht!!!]
\caption{Heliocentric velocity $V_r$ measured in the spectrum of V1648\,Aql obtained 
        on August 7, 2017.  The number of measured lines of each type is given in parentheses}
\begin{tabular}{c|c|c|c} 
\hline
Lines or an element &\multicolumn{3}{c}{V$_r$,  km~s$^{-1}$  } \\ 
\cline{2-4} &stellar atmosphere &CS &IS \\ 
\hline
Absorptions &$10.18\pm0.05~(349)$ &&\\ Emissions
&$8.44\pm0.28~(18)$ &&\\ H$\alpha$ (core) &$9.2$ &&\\ H$\beta$
(core) &$10.6$ &&\\ Na\,I &$9.2~(2)$ &$-3.4~(2)$ &$-12.8~(2)$
\\ K\,I &&$-3.7~(1)$ &\\ DIBs &&&$-12.5\pm0.2~(20)$ \\ 
\hline
\end{tabular} 
\label{velocity} 
\end{table*}

One-dimensional data were extracted from the 2D echelle spectra with the modified (taking 
into account the features of the echelle frames of the spectrograph used) ESO MIDAS reduction 
system ECHELLE context (see the details in paper~[15]). Cosmic ray traces removal was made 
by the median averaging of two spectra successively obtained one after the other. Wavelength
calibration was carried out using spectra of the Th--Ar hollow-cathode lamp. Further reduction 
including the photometric and positional measurements was performed with the latest version
of the DECH20t code~[16]. Note that this program code we traditionally use to reduce the 
spectra allows us to measure the radial velocities for separate features of line profiles. 
Systematic heliocentric velocity V$_r$ measurement errors, estimated from sharp interstellar 
components of Na\,I lines do not exceed 0.25~km~s$^{-1}$ (from a single line), mean
random errors for shallow absorptions are of about 0.5\,km~s$^{-1}$~per line. Thus, for our 
average values in Table~\ref{velocity}, the random errors are about 0.2\,km s$^{-1}$. We have 
identified the lines in the spectrum of V1648\,Aql using the atlas (published earlier in~[17]) 
of the optical spectrum of a canonical post-AGB star HD\,56126 (IRAS\,07134+1005, Sp=F5\,Iab)
based on the data observed at the 6-m  telescope with the same NES spectrograph.

\section{Main results}\label{results}

\subsection{Peculiarities of the V1648\,Aql spectrum  and  radial velocity pattern}

\begin{figure}[bpt!!!]
\includegraphics[angle=0,width=0.5\textwidth, bb=20 70 545 675,clip ]{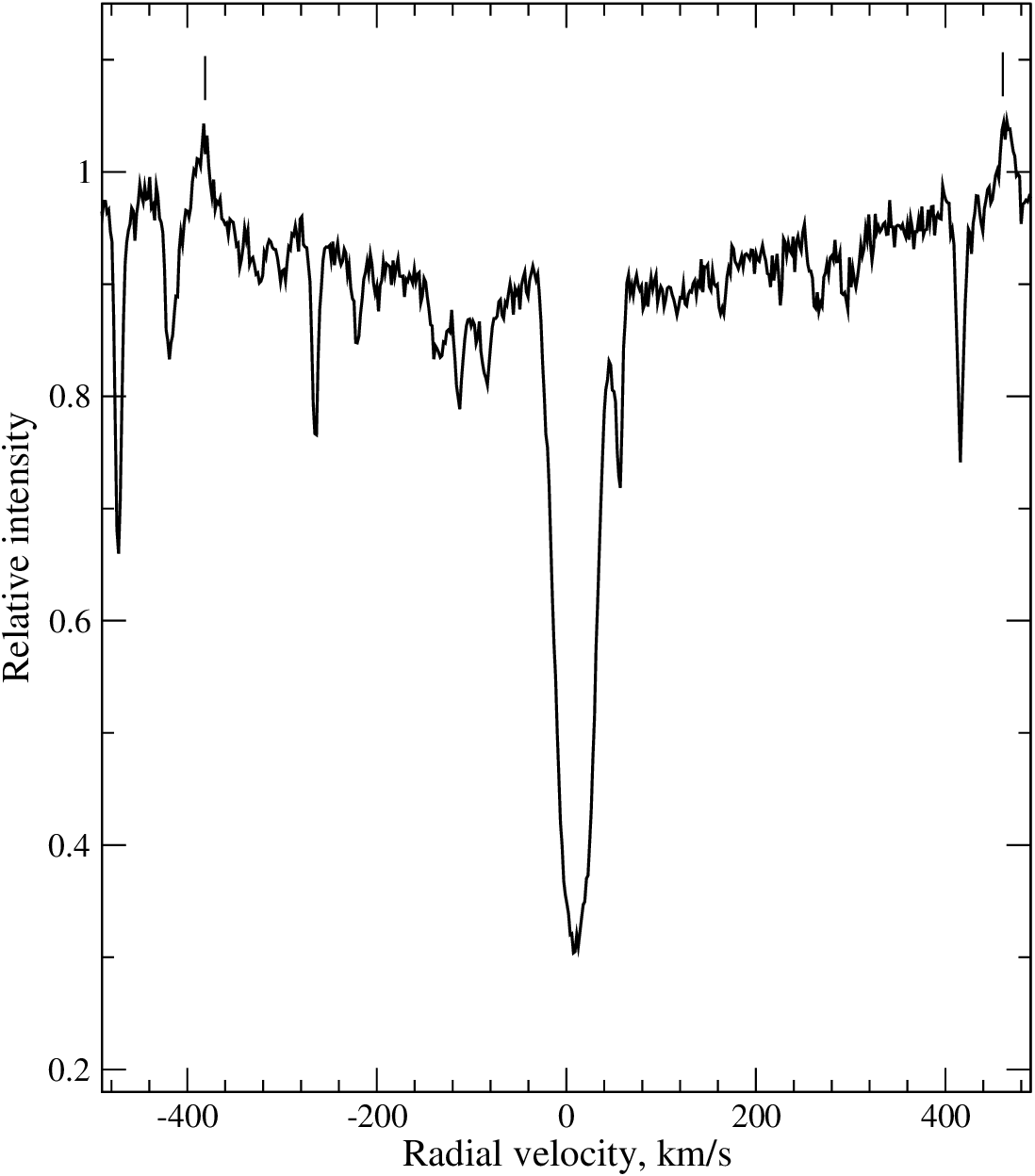}
\caption{H$\alpha$ profile in the spectrum of V1648\,Aql. The vertical lines show the positions of 
         Ti\,I\,6554.23~\AA\ and  Ca\,I\,6572.80~\AA\ emissions.} 
\label{Halpha}
\end{figure}

The optical spectrum of V1648\,Aql in general corresponds to the expected  
spectrum of an F5I supergiant meanwhile possessing several  peculiarities. 
Firstly, the H$\alpha$ profile given in Fig.\,\ref{Halpha} in the relative
intensity--radial velocity coordinates is complex and includes broad wings
and a narrow core indicating a structured atmosphere of the  supergiant with 
an envelope. Given that the H$\beta$ profile is purely absorption without any 
visible emission details. As it follows from Table~\ref{velocity}, the position 
of the H$\beta$ core coincides with the position of atmospheric absorptions of metals. 
The H$\alpha$ profile in the spectrum of V1648\,Aql obtained by the authors of~[13] 
in 2000~does not differ from that we obtained almost 17~years on. This profile 
type is typical of post-AGB stars; Fig.\,6 from~[18] with the profiles in the spectra of four 
southern post-AGB stars can serve as an example of this.

In the spectrum of the already mentioned post-AGB-related star V887\,Her, the H$\alpha$ 
profile demonstrated in Fig.\,2 of~[19] also comprises similar components. Moreover, 
a narrow core of this profile has envelope emissions which are less prominent  in the spectrum 
of V1648\,Aql. The same H$\alpha$-profile type appears at certain moments of observations
of HD\,56126. Figure~1 in the atlas~[17] gives a useful comparison of H$\alpha$ profiles 
in the spectrum of HD\,56126 and in the spectrum of a classical massive supergiant $\alpha$\,Per.

\begin{figure*}[bpt!]
\includegraphics[angle=0,width=0.7\textwidth, bb=40 45 705 520,clip]{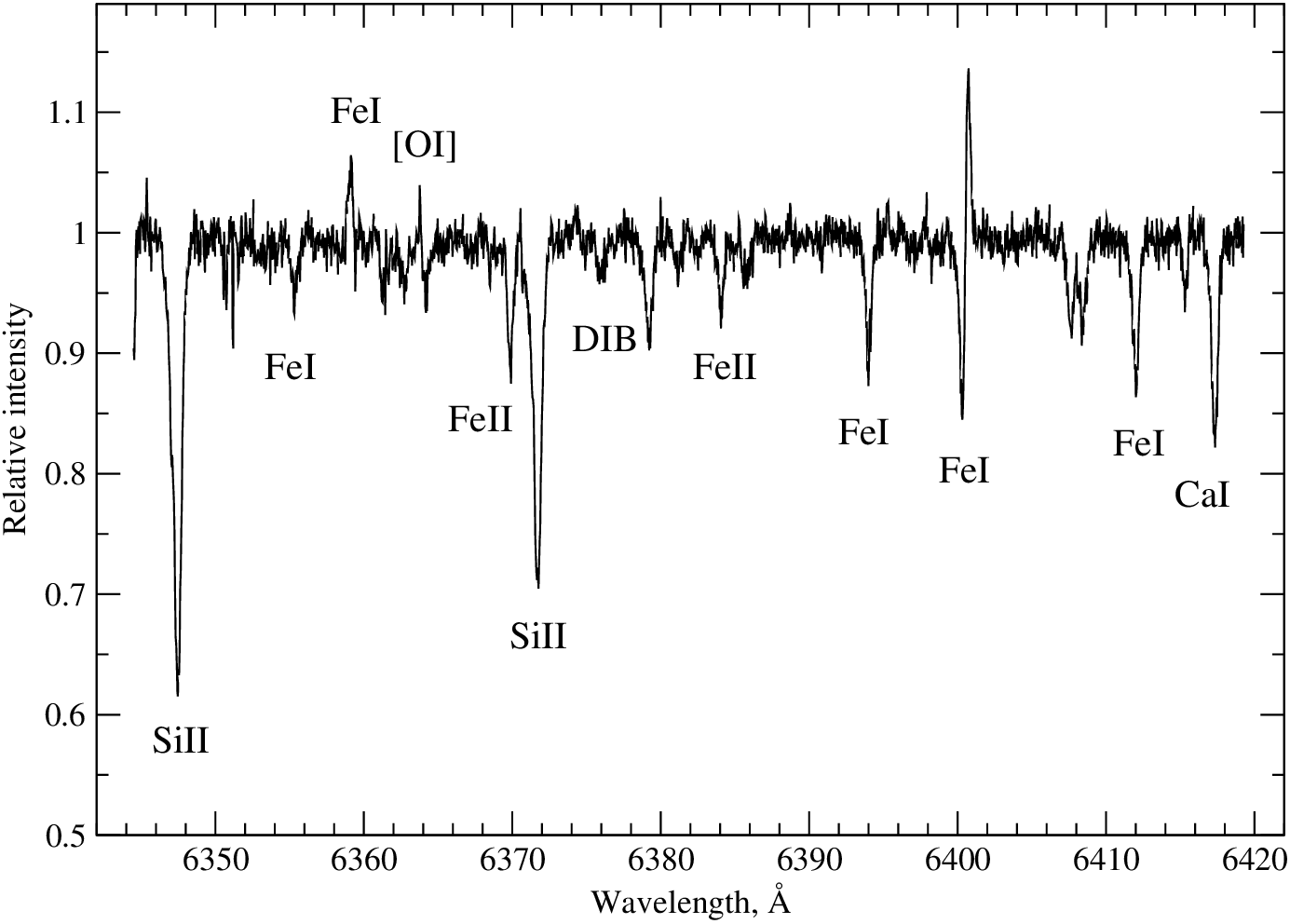}
\caption{A fragment of the V1648\,Aql spectrum containing strong Si\,II(2)\,6347 and 6371\,\AA\ 
   absorptions, Fe\,I\,6359 and 6400\,\AA\ emissions, an interstellar feature (DIB) $\lambda=6376$\,\AA\, 
   and an [O\,I]~6363\,\AA\ emission of ionospheric origin.  The identifications of main features of 
   the fragment are marked.} 
\label{fragm} 
\end{figure*}

The average velocity V$_r$=10.18\,km~s$^{-1}$ (V$_{\rm LSR}$=18.1\,km s$^{-1}$)
we have obtained from numerous metal absorptions in the spectrum of V1648\,Aql proved to be in
agreement with the velocity profile from  OH maser bands for the associated IRAS\,19386+0155 
source~[3] which allows us to accept the value V$_{\rm LSR}$=18.1\,km s$^{-1}$ as the systemic 
velocity of V1648\,Aql. Note that the OH profiles for the IRAS\,19386+0155 source have a large 
width of $\Delta V_r\approx50$~km~s$^{-1}$~[3] which does not allow us to  determine  
more accurately the value of LSR only from the radio data.  Let us consider below the radial 
velocity pattern  from special groups of  spectral features identified for the first time in 
the V1648\,Aql spectrum.

\subsubsection{Metal emissions}
Spectral fragment in Fig.\,\ref{fragm} illustrates the following feature of the optical spectrum 
of V1648\,Aql: the presence of weak emissions of neutral metals with a low excitation
potential of the lower energy level. Two similar emissions in the H$\alpha$ wings are clearly 
seen in Fig.\,\ref{Halpha}. Table~\ref{emis} lists all the emissions of such kind we identified 
in the registered wavelength range. The last column of the table shows the velocity values 
corresponding to the emission positions. The average velocity over 18 emission
features, V$_r{\rm (emis)}$=8.44$\pm$0.28\,km\,s$^{-1}$, is slightly different from the average 
velocity obtained from absorptions, V$_r$=10.18$\pm$0.05\,km s$^{-1}$. However, taking into
account a high accuracy of the average values, we can suspect a weak velocity gradient in
the stellar atmosphere. Halfwidths of these emissions are about 0.3\,\AA, or 
$\Delta$V$_r\approx$13\,km\,s$^{-1}$,   exceeding the halfwidths of the forbidden ionospheric [OI] 
emissions in the spectrum  by 2--2.2 times, what confirms the formation of the emissions more likely 
in the atmosphere of V1648\,Aql. Note that  Fig.\,1 from~[13] also reveals emissions in  the H$\alpha$ 
wings, however, the authors disregarded this feature.

\begin{table*}[ht!]
\caption{A list of metal emissions in the V1648\,Aql spectrum}
\begin{tabular}{r c | r }
\hline
\small $\lambda$\,(\AA{}) & \small Element  &\small  V$_r$, km\,s$^{-1}$ \\ 
 \hline 
 5644.14  &TiI  & 11.20   \\ 
 5847.00  &CoI  &  8.00   \\ 
 5956.70  &FeI  &  8.78   \\ 
 6007.31  &NiI  &  9.83   \\ 
 6108.11  &NiI  & 10.96   \\ 
 6191.19  &NiI  &  6.34   \\ 
 6280.62  &FeI  &  6.57   \\ 
 6358.69  &FeI  &  9.75   \\ 
 6498.95  &FeI  &  8.08   \\ 
 6554.23  &TiI  &  8.20   \\ 
 6572.80  &CaI  &  8.41   \\ 
 6574.24  &FeI  &  7.13   \\ 
 6624.84  &VI   & 10.54   \\ 
 6743.12  &TiI  &  9.17   \\ 
 7052.87  &CoI  &  8.93   \\ 
 7138.91  &TiI  &  7.21   \\ 
 7357.74  &TiI  &  9.05   \\ 
 7714.31  &NiI  &  9.67   \\ 
\hline                 
\end{tabular}   
\label{emis}
\end{table*}

Such less excited emissions of neutral metals were detected earlier~[20] in 
the spectrum of a post-AGB candidate LN\,Hya (IRAS\,12538$-$2611) having an F3\,Ia
spectral type similar to that of V1648\,Aql. A part of the emission details of the 
above-mentioned type from Table\,\ref{emis} is present in the LN\,Hya spectrum
too. In the case of LN\,Hya, the metal emissions appeared in the spectra obtained 
at the times of observations at its active phases in 2010, when a reverse P\,Cyg-type 
profile differed significantly from the profile observed at the  quiet phases. Along 
with this, the position of the H$\alpha$ absorption component also differed
considerably from that at other observation times. Moreover, the H$\alpha$ core was notably 
(by about 15\,km\,s$^{-1}$) shifted towards the long-wave region relative to symmetric metal
absorptions.

Some    Fe, Co, and Ni emissions from Table\,\ref{emis} are also observed in the spectrum of 
a yellow hypergiant $\rho$\,Cas with an extended envelope (see paper~[21] for details 
and required references). Moreover, in the $\rho$\,Cas  spectrum, the average velocity from 
these emissions  inconsiderably varies with time and differs little from the systemic velocity 
of the hypergiant. A small width of these emissions in the $\rho$\,Cas spectrum and an
agreement of the velocity with the systemic velocity are indicative of the fact that these 
weak emissions form in the outer extended gaseous envelope, whose dimensions significantly 
exceed the photometric radius of the star. Emission lines are mainly observed in the periods 
when the stellar brightness decreases, which can be indicative of the relative stability of 
the emission intensity observed at the background of the  weakened photosphere spectrum.

\begin{figure}[bpt!!!]
\includegraphics[angle=0,width=0.6\textwidth, bb=20 70 545 675,clip]{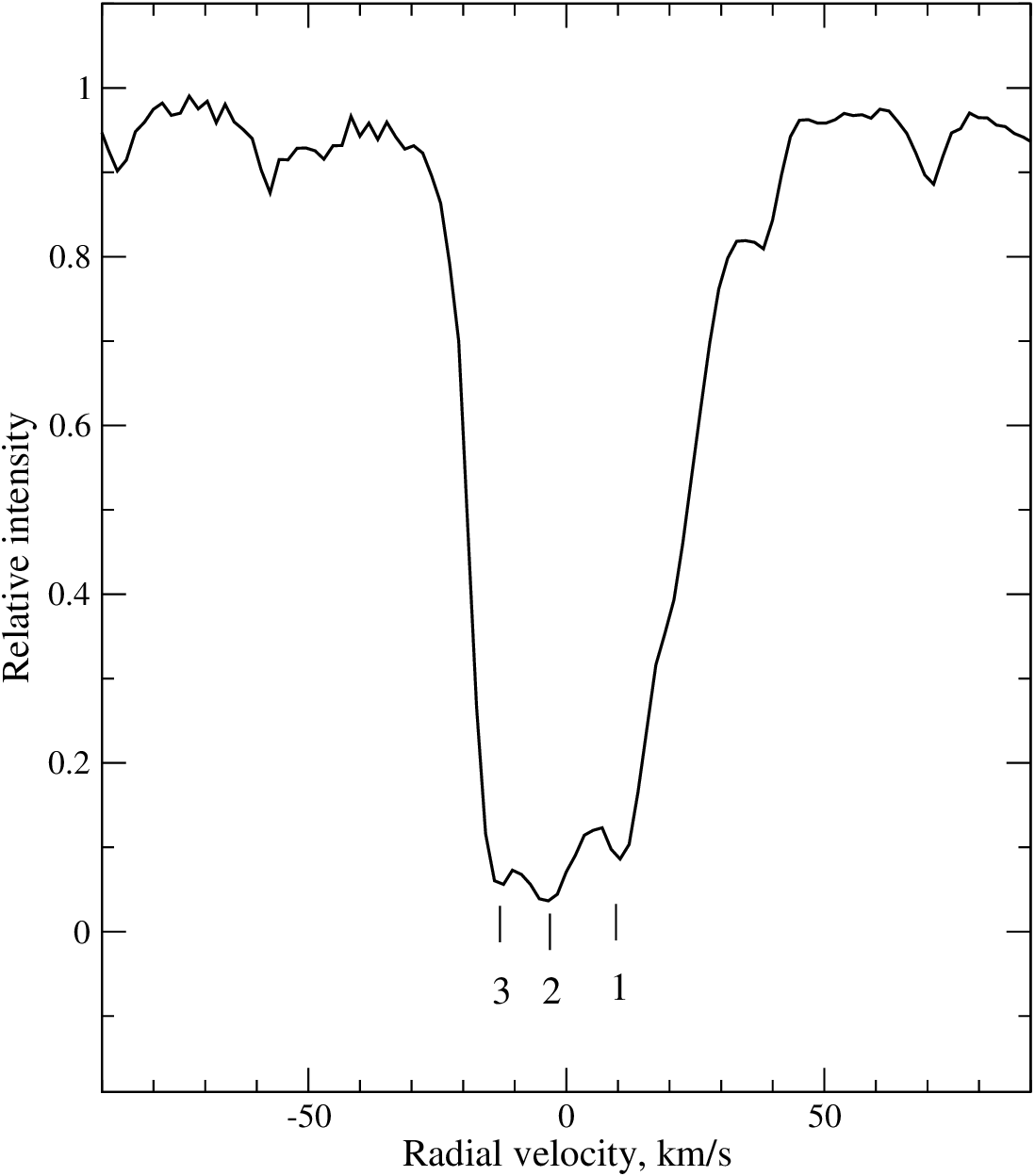}
\caption{The NaI\,5895\,\AA\ line profile in the  relative intensity--radial 
  velocity coordinates. The following positions of the profile components are marked: 
   {\it 1} -- forms in the stellar atmosphere; {\it 2} -- forms in the circumstellar envelope, 
    {\it 3} -- forms in the interstellar medium. }
\label{NaD} 
\end{figure}

\subsubsection{DIBs and a multicomponent profile of Na\,I D lines}

The optical spectrum of V1648\,Aql contains  numerous interstellar features despite a considerable 
distance to the Galaxy plane (the galactic latitude of the star is $|b|>10\degr$).
Table\,\ref{DIBs} presents diffuse interstellar bands (DIBs) from the well-known list of 
Jenniskens and D\'esert~[22]. We have identified and reliably differentiated these DIBs among 
the blends in the V1648\,Aql spectrum. For these features, the table lists radial velocities 
corresponding to band positions and their equivalent widths W$_\lambda$. The measured 
equivalent widths for several common DIBs agree well with the data measured in the V1648\,Aql 
spectrum by the authors of~[23]. However, radial velocities do not agree.

The  high-quality spectrum allowed us to resolve into components the NaI\,5889 and 5895\,\AA\ 
D-lines    and measure the position of the interstellar  KI\,7696\,\AA\ absorption   for the
first time ever. The NaI\,5889 and 5895\,\AA\ line profiles in the V1648\,Aql spectrum 
confidently reveal individual components, the averaged position of which is given in
Table\,\ref{velocity} and in Fig.\,\ref{NaD}. The position of the long-wavelength
component, V$_r$=9.2\,km\,s$^{-1}$,  complies with the average velocity V$_r$ measured from a 
large selection of metal absorptions within the above error. This testifies the formation of 
this component in the stellar atmosphere. The location of the shortest-wavelength component of 
the Na\,I doublet lines, V$_r=-12.8$\,km\,s$^{-1}$, coincides with the average velocity 
V$_r$(DIBs)=$-12.5\pm0.2$\,km~s$^{-1}$  obtained from the DIBs set identified in the spectrum, 
which allows us to confirm that this  component is formed in the interstellar medium. 
Figure\,\ref{NaD} shows a differing steepness of wings of the atmospheric and interstellar 
components, which also proves our version on their formation regions.

The component of the NaI D-lines with the velocity V$_r$=$-3.4$\,km\,s$^{-1}$ is shifted towards the
short-wavelength region by 13.6\,km\,s$^{-1}$ relative to the average radial velocity from the 
atmospheric absorptions. We can naturally suppose that this component forms in the circumstellar
envelope expanding with the velocity V$_{\rm exp}$=13.6\,km\,s$^{-1}$ typical of post-AGB stars 
(see numerous examples in~[24, 25] for comparison).

\begin{table*}[ht!]
\caption{DIBs parameters in the V1648\,Aql spectrum}
\begin{tabular}{ c|  c   r }
\hline
\small $\lambda$, \AA{}  &\hspace{3mm} \small  V$_r$, km\,s$^{-1}$ &  \hspace{2mm} W$_{\lambda}$, m\AA{} \\ 
 \hline 
 5456.00 &   $-13.73$  &  15   \\  
 5487.67 &   $-12.99$  &  18   \\  
 5512.68 &   $-14.02$  &  16   \\  
 5780.48 &   $-12.43$  &  200  \\  
 5849.81 &   $-11.65$  &  7    \\  
 5910.57 &   $-14.10$  &  12   \\  
 6089.85 &   $-10.09$  &  13   \\  
 6158.57 &   $-11.29$  &  41:  \\  
 6195.98 &   $-14.14$  &  40   \\  
 6203.05 &   $-12.60$  &  88   \\  
 6234.03 &   $-11.16$  &  22   \\  
 6269.85 &   $-13.62$  &  33   \\  
 6376.08 &   $-13.63$  &  19   \\  
 6379.32 &   $-13.64$  &  55   \\  
 6445.28 &   $-11.92$  &  6    \\  
 6449.22 &   $-14.83$  &  7    \\  
 6613.62 &   $-12.37$  & 144   \\  
 6660.71 &   $-12.23$  &  28   \\  
 7367.13 &   $-12.76$  &  28   \\  
 7651.40 &   $-13.22$  &  15   \\  
\hline                
\end{tabular}   
\label{DIBs}
\end{table*}

\subsubsection{Luminosity and distance to the star}

The Gaia\,DR2 catalog gives an unreliable (negative) parallax for V1648\,Aql (as well 
as for some other distant stars with extended dust envelopes) which already indicates 
a vast distance to the object. Let us try to estimate the luminosity of the star and the distance
to it based on the spectroscopic data we obtained. To estimate the E(B$-$V) color excesses 
determined by the interstellar extinction, we use the measured intensities for the selected DIBs 
and the calibrations E(B$-$V)$\leftrightarrow$W$_\lambda$(DIBs) according to the data from~[26]. 
Using only  eight mostly reliably measured bands with W$_\lambda>$10\,m\AA\ from Table~\ref{DIBs}, 
we obtain the average color excess of E(B$-$V)=$0\fm68$. This color excess index is caused by
the interstellar extinction only. Some fraction of the complete reddening of the star is caused  
by the extinction not in the interstellar but rather in the circumstellar medium. Thus, it is
natural that our  color excess estimate is somewhat smaller than the value of E(B$-$V)=$0\fm8-0\fm9$
in~[10].

We will estimate the stellar luminosity using the well-known luminosity criterion for the evolved 
stars --- the intensity of the oxygen triplet OI\,7774\,\AA{}. The total equivalent width of the
triplet in our spectrum equals W$_\lambda$(7774)=1.42\,\AA{} which is typical of post-AGB stars 
(see the data in [27] for comparison).  Along with this, the equivalent width of the triplet in 
the V1648\,Aql spectrum is 1.5$\div$2~times lower than the equivalent width of the triplet
in the spectra of high-luminosity massive stars of a close spectral type, V1302\,Aql~[28] and 
V509\,Cas~[29]. Taking into account the calibration of M$_V\leftrightarrow$W$_\lambda$(7774),
according to~[30], we obtain the V1648\,Aql luminosity corresponding to  the intensity of the 
triplet in the spectrum: M$_V\approx-5^{\rm m}$.  Results of Takeda et al.~[30] allow us to notice 
a good agreement of two spectral  parameters obtained from the V1648\,Aql spectrum, namely, 
the surface gravity and W$_\lambda$(7774).

The luminosity we obtained agrees well with the spectral classification and the luminosity expected 
from theoretical concepts on the evolution of post-AGB stars~[31]. However, the luminosity
estimate obtained from W$_\lambda$(7774) is not quite accurate due to a number of reasons. 
Firstly, V1648\,Aql shows a long-term trend of the apparent brightness at about $0\fm4$~[10]. Secondly, 
the equivalent width of the  OI\,7774~\AA{} triplet can be increased due to a great oxygen overabundance  
in the atmosphere of V1648\,Aql. Kovtyukh et al.~[32] considered the influence of stellar parameters, 
including their  metallicity on the calibration  M$_V\leftrightarrow W_\lambda$(7774). 
Along with this, they justly noted an inconsiderable effect of the oxygen abundance on 
W$_\lambda$ of the strong lines of the OI\,7774\,\AA{} triplet.

Taking into account the apparent brightness trend from paper~[10] and the absorption value
A$_V$=$2\fm17$ (with the standard value of R=3.2), we obtain the distance to the star of d$\approx$5\,kpc. 
If we also consider the color excess due to absorption in the stellar envelope which is about $0\fm4$ 
according to Arkhipova~et~al.~[10], then the distance to the star will decrease to 3.8\,kpc 
(at R=3.2) and even to 1.8\,kpc (at R=7.4). Here we should mention the distance
estimate to IRAS\,19386+0155 in the catalog~[33]. Modeling the SED, these authors came to the 
luminosity estimate L/L$_{\sun}$=6000 which is close to our luminosity determination from the oxygen 
triplet.  In this regard, they derived a small color excess E(B$-$V)=$0\fm36$
and the distance d=3.3~kpc. These estimates are precarious, if we take into account an anomalous 
character of the SED for IRAS\,19386+0155, conditioned by  the presence of warm (of about
1000\,K) and cold (of about 200\,K) dust~[13].

On the whole, we have to admit that the distance to V1648\,Aql has not so far been determined precisely, 
although we believe its bottom estimate to be d$\approx$1.8~kpc.

\subsection{Chemical composition of the V1648\,Aql atmosphere}

While  we estimate the  chemical composition of the stellar atmosphere, it is always quite  complicated  
to state the main parameters: the effective temperature T$_{\rm eff}$ and the surface gravity $\log g$. 
The task becomes even more complex for an object with a circumstellar dust envelope, for which it is 
difficult to use the photometric data in determination of an effective temperature due to an uncertain 
reddening. As initial parameters for the model T$_{\rm eff}$, $\log g$ and microturbulent velocity $\xi_t$,
we used the values determined for V1648\,Aql with a high accuracy within the purely spectral approach~[13].
These authors obtained reliable chemical element abundances based on a high-quality spectrum using a 
modern analysis method. We shall therefore consider below the chemical composition that we have
obtained only in brief, highlighting some key points.

\begin{table*}[bpt!!!]
\caption{Chemical element abundances $\log \varepsilon$(X) in the V1648\,Aql atmosphere. 
   The error of mean abundance $\sigma$ obtained from a number of lines $n$ is listed. 
   The chemical composition of the solar atmosphere is taken from~[37]. 
   The last column gives the relative abundances in the atmosphere of a related star 
   V887\,Her~[38].} 
\begin{tabular}{l|c|l|c|c|c|c|c} \hline
\multicolumn{2}{c|}{The Sun} &\multicolumn{5}{c|}{V1648\,Aql~(the present paper)} &V887\,Her~[38]\\ 
\hline
Elm &$\log\varepsilon$\,(E) &\multicolumn{1}{c|}{X}
&$\log\varepsilon$\,(X) &$\sigma$ &$n$ &${\rm [X/Fe]}_{\sun}$
&${\rm [X/Fe]}_{\sun}$ \\ 
\hline
C &8.39 &C\,I &8.47 &0.10 &13 &$+0.75$ &$+0.43$ \\ N &7.86 &N\,I
&7.88 &&1 &$+0.69$ &$\le+0.46$ \\ O &8.73 &O\,I &9.42 &0.26 &3
&$+1.36$ &$+0.77$ \\ Na &6.30 &Na\,I &6.01 &0.16 &4 &$+0.38$
&$+0.79$ \\ Mg &7.54 &Mg\,I &7.43 &0.16 &3 &$+0.48$ &$+0.45$ \\ Si
&7.52 &Si\,I &7.05 &0.17 &7 &$+0.20$ &$+0.58$ \\ &&Si\,II &7.41
&0.14 &2 &$+0.56$ &$+0.10$ \\ S &7.14 &S\,I &7.55 &0.30 &4
&$+1.08$ &$+0.77$ \\ Ca &6.33 &Ca\,I &5.87 &0.14 &13 &$+0.21$
&$+0.05$ \\ Sc &3.07 &Sc\,II &2.47 &0.15 &9 &$+0.07$ &$-0.16$ \\
Ti &4.90 &Ti\,I &4.26 &0.45 &3 &$+0.03$ &$-0.07$ \\ &&Ti\,II &4.26
&0.14 &12 &$+0.03$ &$0.00$ \\ Cr &5.64 &Cr\,I &5.12 &0.16 &9
&$+0.14$ &$-0.02$ \\ &&Cr\,II &5.14 &0.09 &17 &$+0.17$ &$-0.19$ \\
Fe &7.45 &Fe\,I &6.77 &0.04 &121 &$-0.01$ &$0.00$ \\ &&Fe\,II
&6.80 &0.09 &26 &$+0.02$ &$0.00$ \\ Ni &6.23 &Ni\,I &5.61 &0.13 &9
&$+0.05$ &$+0.17$ \\ Zn &4.62 &Zn\,I &4.02 &0.27 &3 &$+0.17$
&$+0.21$ \\ Y &2.21 &YI\,I &1.15 &0.25 &4 &$-0.40$ &$-0.58$ \\ Ba
&2.17 &Ba\,II &1.39 &0.25 &4 &$-0.11$ &$-0.22$ \\ Ce &1.61 &Ce\,II
&0.79 &&1 &$-0.15$ &\\ Eu &0.52 &Eu\,II &0.35 &0.10 &3 &$+0.50$
&$+0.34$ \\ 
\hline
\end{tabular} 
\label{chem} 
\end{table*}

Endeavoring to keep the ionization balance for Fe\,I and Fe\,II,
we performed many calculations of their abundances varying the
main parameters of the atmosphere model. As final values, we accepted the following: 
T$_{\rm eff}$=6800$\pm$100~K, $\log g$=1.2$\pm0.2$,  $\xi_t$=8.3$\pm$0.5~km\,s$^{-1}$ which 
within the error agree with the parameters from~[13]. The  choice of the effective temperature 
and surface gravity is confirmed by a good agreement of abundances from absorptions of neutral 
atoms and ions for Ti and Cr. Herein, the agreement is worse for silicon which is of little 
significance owing to a small number of  SiII lines. When determining the parameters of the model 
atmosphere and calculating the abundances, we used the lines of small and moderate intensity
with the equivalent widths of W$_{\lambda}\le0.25$\,\AA{}, since the approximation of the 
stationary plane-parallel atmosphere can be inadequate while describing the stronger spectral features. 
All the absorption equivalent widths are measured in the Gaussian approximation. Oscillator strengths 
$\log gf$ and other atomic constants are adopted from the VALD database~[34, 35]. We performed the
calculations of plane-parallel models and chemical abundance in the approximation of the local 
thermodynamic equilibrium (LTE) using the latest version of the software designed and adapted for
the OS~Linux environment by V.\,Tsymbal~[36].

Table~\ref{chem} gives the obtained average element abundances  $\log \varepsilon$(E), as well 
as the relative abundances ${\rm [X/Fe]}_{\sun}$.  The chemical composition of the solar photosphere 
given in the second column,  relative to which we consider the abundances of chemical elements of 
the star studied, is borrowed from~[37]. The scatter of the chemical element abundances obtained from 
a set  of lines is insignificant: the mean error ${\sigma}$ generally does not exceed 0.3\,dex for 
the elements with more than four absorptions used (see Table\,\ref{chem}).

\subsubsection{Iron-peak elements}

The iron abundance, usually accepted as stellar metallicity, in the atmosphere of V1648\,Aql differs 
from the solar abundance: $\lg\varepsilon$(FeI)=6.82. The reliably determined abundances of titanium, 
chromium and nickel belonging to the iron group also differ little from normal: (Ti,Cr\,I,Ni)/Fe$_{\sun}$=+0.12.

\subsubsection{Light elements}
When analyzing the characteristics of the chemical abundance of the V1648\,Aql atmosphere, one should 
consider the results of modeling its energy distribution (SED)~[13], based on which the envelope 
of the star is referred to the O-rich type. According to current concepts~[7, 39], the presence of 
freshly generated lithium, whose synthesis is due to the HBB-process, can be expected in the atmosphere of
an O-rich star at the post-AGB stage. The LiI\,6707.8\,\AA{} line is absent in the spectrum of V1648\,Aql, 
which is indicative of its belonging to stars with initial masses below 4\,M$_{\sun}$~[40].

A considerable oxygen excess  is detected in the atmosphere: its relative abundance is 
[O/Fe]$_{\sun}$=+1.36$\pm0.26$. Given  a smaller carbon excess [C/Fe]$_{\sun}$=+0.75$\pm0.10$, 
we obtain  a ratio  ${\rm O/C}>1$. A significant excess of oxygen is consistent with  the fact that 
V1648\,Aql is a low-mass supergiant. An analysis of the spectra of massive stars [41] shows that, 
in accordance with the theoretical predictions,  the evolution of massive stars results in the 
deficiency of oxygen, which is converted into nitrogen during the CNO cycle. Therefore, nitrogen 
is an essential element for reliable determination of the evolutionary status of a star. In the 
registered spectral range, only one of its lines is available to us: NI\,7468\,\AA{}. 
The nitrogen abundance, [N/Fe]$_{\sun}$=+0.69, is calculated based on the measured equivalent width 
of this rather weak line, W$_{\lambda}$=22\,m\AA{}. The ratio ${\rm O/N}>{\rm 1}$ also confirms the 
low-mass supergiant status of V1648\,Aql.

We determined the sodium abundance from the weak subordinate NaI\,5682, 5688, 6154, and 6160\,\AA{} 
lines, for which the corrections  caused by a shift  from the LTE are minimal~[42]. The detected sodium excess,
[Na\,I/Fe]$_{\sun}$=+0.38$\pm 0.16$, is small and does not exceed $3\sigma$. However, taking into 
consideration the abundances of other metals of the $\alpha$-process (Mg, Si, S, and Ca), we can speak 
about the presence of their small excess that is typical of unevolved stars with the metallicity of 
[Fe/H]$_{\sun}< -0.5$~[43].

\subsubsection{On selective depletion of chemical elements}

For stars with gas and dust envelopes, an effective mechanism that creates chemical abundance anomalies 
can be the selective separation of chemical elements. In the star under study with an IR flux excess, an 
intensive exchange of matter of the atmosphere and the circumstellar gas-and-dust envelope can occur, as
V1648\,Aql has an effective temperature suitable for this process~[44].  In general, the analysis of 
chemical element abundances is complicated by the fact that the abundances of individual elements 
(CNO triad, heavy metals) can be influenced   both by the nuclear processes during the stellar
evolution and by a selective depletion of atoms on the dust particles. A possible shift from the LTE 
for all chemical elements given in Table\,\ref{chem} does not exceed 0.1\,dex (see paper~[44] and 
references therein).

\begin{figure}[bpt!!!] 
\includegraphics[angle=0,width=0.5\textwidth]{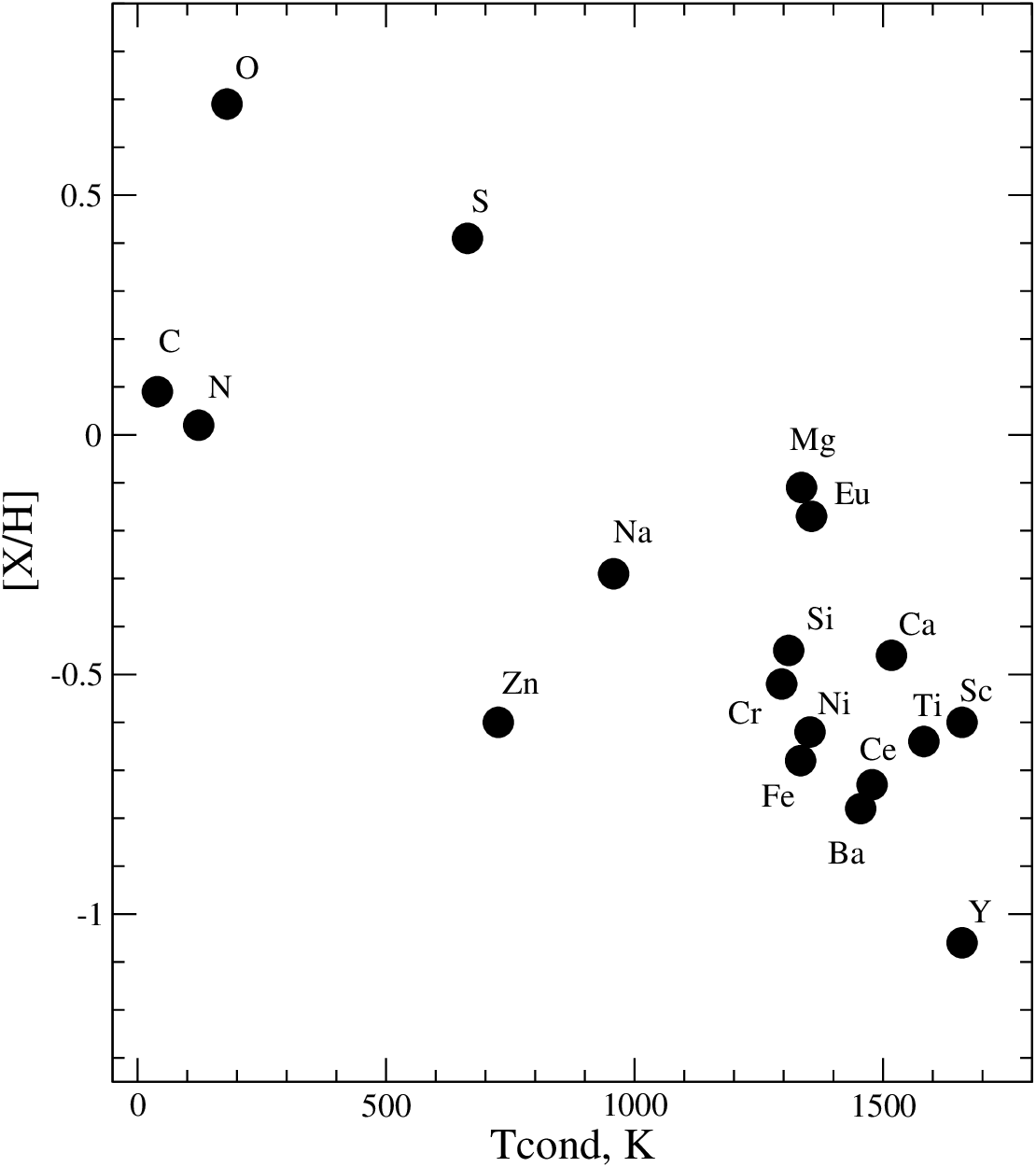}
\caption{Abundances of chemical elements in relation to the solar ones [X/H]$_{\sun}$ in the atmosphere 
  of V1648\,Aql depending on the condensation temperature T$_{\rm cond}$  from~[37]. } 
\label{Depletion}
\end{figure}

The  dependence of the element abundances we obtained [X/H] from Table~\ref{chem} on  the condensation 
temperature T$_{\rm cond}$ from~[37] in Fig.\,\ref{Depletion} indicates the presence of a moderate 
selective depletion in the  V1648\,Aql system. We can expect the same behavior of elements with similar 
values of T$_{\rm cond}$. For example, according to~[37], T$_{\rm cond}$ is similar for calcium and 
scandium, as well as for barium and cerium. Indeed, as follows from  Table\,\ref{chem} and 
Fig.\,\ref{Depletion}, these pairs of elements have similar relative abundances.

Figure~\ref{Depletion} shows a considerable difference of the [X/H] abundances for sulfur and 
zinc with similar values of T$_{\rm cond}$. What is important in this pair of elements is the
zinc abundance, which, being little susceptible to the depletion process, does not change 
its abundance in the course of the nuclear evolution of this star. As it was already shown in 1991 
by Sneden et al.~[45], in a large range of metallicities, the zinc  abundance corresponds to the 
behavior of iron, [Zn/Met]=+0.04.  The conclusion by Sneden et al. refined the results of Mishenina et
al.~[46]: according to this work, zinc corresponds to metallicity in a wide range of its values, 
[Fe/H]=$-0.5 \div -3.0$. In the V1648\,Aql  atmosphere, the relative abundance of zinc is 
[Zn/Fe]=+0.17$\pm0.27$, which indicates a small and statistically insignificant depletion of iron atoms.

The sulfur excess, poorly susceptible to selective condensation is [X/H]=+0.41, and relative to 
iron [X/Fe]$_{\sun}$=+1.08, which is significantly higher than the abundance excesses of other
$\alpha$ process elements. A large difference between the relative abundances  of sulfur and zinc 
is a common phenomenon for post-AGB stars and RV Tau-type stars, which is not explained so far~[44]. 
As Fig.\,3 in~[47] and Figs.\,5 and~6 in a  recent publication~[48] demonstrate, this difference 
can reach an order of  magnitude and greater. As follows from paper~[47], a great [S/Zn] ratio is
typical of stars in the thick disk. The belonging of V1648\,Aql to the population of the thick disk 
is confirmed by the magnitude of the ratio [Zn/H]$_{\sun}$=$-0.6$~[44].

\subsubsection{Heavy metals}
No excess of  the $s$-process heavy metals  (Y, Ba, and Ce) has been detected in the atmosphere of V1648\,Aql. 
Therefore, we can speak about the inefficiency of the third dredge-up. The absence of  the expected excess 
of heavy metals with respect to iron is a fact known for post-AGB supergiants. The deficiency of the 
$s$-process elements in the atmospheres of post-AGB stars is observed much more often than their 
excess~[25, 49--51]. The presence or absence of an excess of the $s$-process elements is associated with 
the initial mass of a star and the mass loss rate at the AGB stage, which determine the evolution of an 
individual star and the mass  of the stellar core~[6]. The deficiency  of heavy metals in the atmosphere of 
the star under study could have been be predicted due to the belonging of IRAS\,19386+0155 to O-rich sources, 
as an excess of $s$-process elements is usually associated with a carbon excess in the atmosphere of 
a star having a carbon-rich envelope~[25]. The estimated excess of europium [X/Fe]$_{\sun}$=0.50 synthesized 
mainly in the rapid neutron capture reactions (the $r$-process) is typical of the atmospheres of post-AGB 
stars (see examples in papers~[18, 50, 51]).

\vspace{0.7cm}

{\bf In whole} we can approve that the parameters we obtained, namely,  luminosity, distance, 
metallicity, and the  features of chemical abundance, agree with the fact that the star is  
at the post-AGB stage in the thick  (or, taking into consideration a small velocity of the star, 
in the old thin) Galaxy disk.

Let us emphasize the lack of variations of stellar effective temperature from 2000 to 2017, which 
could be expected due to the observed increase of the apparent brightness~[10]. Note that the
atmospheric parameters and the main features of the chemical abundances pattern of two 
high-latitude post-AGB stars V1648\,Aql and V887\,Her~[19, 38] compared in Table\,\ref{chem} 
confirm  the similarity of both objects. However, they have fundamentally different distributions of 
energy: unlike V1648\,Aql, V887\,Her has a double-peaked distribution typical of post-AGB stars.
In the parameter domain, similar objects also located outside the Galactic plane are LN\,Hya~[20, 44, 47] 
and the central star of the IR source IRAS\,18025$-$3906~[48].

Based on the features of the anomalous SED, the authors of~[13] drew  another fundamental conclusion: 
a possible presence of a circumstellar dust  disk in the V1648\,Aql system may indicate the presence 
of a component in the system. Considering the anomalous brightness variability of the star too [10, 11], 
spectral monitoring is required in order to search for the variability of the spectrum, the velocity 
field, and to identify a possible binarity of the star.

\section{Conclusions}\label{conclus}

Using the 6-m BTA telescope combined with the NES echelle spectrograph (R$\ge60\,000$), we obtained 
the optical spectrum of V1648\,Aql, the central star of the IRAS\,19386+0155, which allowed us to 
study for the first time the kinematic state of its extended atmosphere, circumstellar envelope, 
and interstellar medium in the direction of the object.

The radial velocity measured from numerous metal absorptions is V$_r$=10.18$\pm$0.05\,km\,s$^{-1}$. 
The velocity averaged over the set of twenty DIBs identified in the spectrum is 
V$_r$(DIBs)=$-12.5\pm$0.2\,km\,s$^{-1}$. 
In the profile of the NaI\,D lines, three separate components are determined:
\begin{itemize}
\item a long-wave component, V$_r$=9.2\,km\,s$^{-1}$, whose position coincides with the
averaged position of metal absorptions, is formed in the atmosphere of  the star; 
\item the most short-wavelength component, V$_r$=$-12.8$\,km\,s$^{-1}$, whose position coincides 
with the average velocity  from a set of twenty DIBs, is formed in the interstellar medium; 
\item the component with the velocity V$_r$=$-3.4$\,km\,s$^{-1}$, shifted towards the short
wavelength region by 12.6\,km\,s$^{-1}$ relative to the average radial velocity from atmospheric 
absorptions, is formed in the circumstellar envelope expanding at a velocity typical of post-AGB stars.
\end{itemize}

We detected narrow emissions in the spectrum with the intensity of about 10\% of the local continuum 
level, which we identified with  low-excited lines of metal atoms. Their position shifted relative to the 
absorptions, with the average value of V$_r$=8.4$\pm0.3$\,km\,s$^{-1}$,  might indicate on the presence of 
a weak velocity gradient in the upper atmosphere of the star.

The intensity of the oxygen triplet OI\,7773~\AA{} corresponds to the luminosity of M$_V\approx-5^{\rm m}$. 
In the absence of reliable parallax of the star, we obtained the bottom distance estimate d$\ge$1.8\,kpc,
taking into account the interstellar and circumstellar reddening.

Using the model atmosphere method, we determined the fundamental parameters and abundances of 18 
chemical elements in the atmosphere confirming the status of a post-AGB star for V1648\,Aql.

\begin{acknowledgements}
The authors thank the Russian Foundation for Basic Research for partial financial support (project 18--02--00029\,a).
In this study we made use of the SIMBAD, SAO/NASA ADS, VALD, and Gaia\,DR2 astronomical databases. 
\end{acknowledgements}


\end{document}